# Machine learning for the recognition of emotion in the speech of couples in psychotherapy using the Stanford Suppes Brain Lab Psychotherapy Dataset

Colleen E. Crangle[a,b], Rui Wang[a], Marcos Perreau-Guimaraes[a], Michelle U. Nguyen[a,c], Duc T. Nguyen[a], Patrick Suppes[a]

[a]*Stanford University, Center for the Study of Language and Information, Ventura Hall, 220 Panama Street, Stanford, CA 94305-4101*

[b]*Converspeech LLC, 60 Kirby Place, Palo Alto, CA 94301-3039*

[c]*A Balanced Approach to Counseling, 25 North 14th Street, San Jose, CA 95112*

**Abstract**

The automatic recognition of emotion in speech can inform our understanding of language, emotion, and the brain. It also has practical application to human-machine interactive systems. This paper examines the recognition of emotion in naturally occurring speech, where there are no constraints on what is said or the emotions expressed. This task is more difficult than that using data collected in scripted, experimentally controlled settings, and fewer results are published. Our data come from couples in psychotherapy. Video and audio recordings were made of three couples (A, B, C) over 18 hour-long therapy sessions. This paper describes the method used to code the audio recordings for the four emotions of Anger, Sadness, Joy and Tension, plus Neutral, also covering our approach to managing the unbalanced samples that a naturally occurring emotional speech dataset produces. Three groups of acoustic features were used in our analysis: filter-bank, frequency, and voice-quality features. The random forests model classified the features. Recognition rates are reported for each individual, the result of the speaker-dependent models that we built. In each case, the best recognition rates were achieved using the filter-bank features alone. For Couple A, these rates were 90% for the female and 87% for the male for the recognition of three emotions plus Neutral. For Couple B, the rates were 84% for the female and 78% for the male for the recognition of all four emotions plus Neutral. For Couple C, a rate of 88% was achieved for the female for the recognition of the four emotions plus Neutral and 95% for the male for three emotions plus Neutral. For pairwise recognition, the rates ranged from 76% to 99% across the three couples. Our results show that couple therapy is a rich context for the study of emotion in naturally occurring speech.

*Keywords:* emotion; emotion recognition, couple psychotherapy; speech; acoustic features; acoustic analysis; machine learning; random forests

---

[1] We would like to acknowledge the collaboration of Cynthia T.M.H. Nguyen, M.D., and Margot Tuckner, Ed.D., MFT, psychotherapists on the team.

Email addresses: crangle@alumni.stanford.edu (Crangle), ruiw2000@gmail.com (Wang), montereyunderwater@gmail.com (Perreau-Guimaraes), michelleun@att.net (Nguyen), dtn006@stanford.edu (Nguyen), psuppes@stanford.edu (Suppes)

# 1. Introduction

Emotional content is an important part of the exchange of information between speaker and listener. It is transmitted through several sensory modalities, including voice and facial expressions. In this paper, we focus on the expression of emotion in speech. Understanding what in the speech waveform codes the emotional content is of interest at a fundamental level, providing insights into how the brain represents emotions. It also has practical application in the design of human-computer interaction systems, as for call centers that use automatic speech recognition and speech synthesis to detect and react to emotional cues from the caller in an emotionally appropriate way.

Our overall interest here is in recognizing emotions in naturally occurring speech, specifically in the verbal interaction between people. Our data come from couples in psychotherapy where the speech was not controlled in any way. Work on the automatic recognition of emotion in speech collected in scripted, experimentally controlled settings has been underway since the 1990s and a number of emotional-speech datasets are now available in several languages. But very little has been done using naturally occurring speech. Not only is the recognition task more difficult in such settings, the constraints of producing recordings of sufficient quality narrow the possibilities of collecting such data. For these reasons, recognition of emotions in naturally occurring speech is much harder, with fewer results having been published and much lower recognition rates achieved.

Ultimately, however, the study of emotions in natural settings is needed. It allows for a deeper understanding of the finer nuances carried by the speech signal. And it has immediate application to practical human-computer systems in that the user typically cannot be constrained to speak only select words, especially when the goal is a more natural form of interaction.

For this study we use data from the Stanford Suppes Brain Lab Psychotherapy EEG Dataset (Suppes et al., 2015). From this database of speech, video and EEG-brain recordings, we used 18 sessions of approximately an hour each for three couples (A, B, and C) undergoing therapy over a period of 2 years. The couples were all male-female couples. We interfered as little as possible with the therapy, only instructing the participants to avoid moving beyond the reach of the recording apparatus. In this paper we analyze the speech recordings of the couples participating in therapy. The videos were used in the coding of emotions and the EEG data were reserved for later analysis. The four emotions of interest in this study were Anger, Sadness, Joy, and Tension - plus Neutral, which is defined as the absence of those four emotions.

An overview of our approach to emotion recognition in naturally occurring speech is as follows. An emotion one out of a designated set of emotions is identified with each unit of language (word or phrase or utterance) that was spoken, with the precise start of each such unit determined in the continuous acoustic signal. Using these start points, equal-length segments of the acoustic signal are demarcated, producing a set of emotion-coded tokens. With a sufficient number of acoustic-signal segments coded for emotions in this way, it is possible to use machine learning to detect what, in the acoustic signal, differentiates the times an utterance is spoken when one emotion is being expressed as opposed to another. The extent to which the emotions are successfully recognized corresponds to how successfully the acoustic-signal segments are classified by a machine learning algorithm as belonging to one or another of the emotions.

There are two distinct challenges to emotion recognition in naturally occurring speech that we address in this paper. The first is that there are no constraints on the words and phrases spoken nor how often they occur. In a controlled setting, typically produced using actors and scripts, the collected speech is limited to a small set of words and phrases with the utterances repeated many times. Both the variety and the number of utterances are controlled. Furthermore, with emotions that are deliberately evoked, the number of emotions and the timing of their appearance with the utterances are controlled. The result is usually balanced datasets with equal numbers of utterances for each emotion, and a limited number of emotions appearing in the dataset. Unbalanced datasets present two kinds of difficulties, which we discuss in detail in Section 6.

The second challenge lies in deciding on the unit of analysis – whether word, phrase or whole utterance, for example, or automatically computed segments of the acoustic signal – and then for each unit determining where in the continuous acoustic stream to start and end each data sample, that is, the time span of each observation. In this study we chose words to provide the precise starting points for our units of analysis. The precise onset of a word was able to be determined with some confidence using the transcription. But there are marked differences in the time taken to speak different words – "I" compared to "afterwards," for example – or even the same word on different occasions, particularly when different emotions are being expressed. We sought to capture enough of each word and not too much of the word that follows. In some instances, part of a word's acoustic signal may be truncated and in others, part of the word that follows will be included. How we determined the end points for our data samples is discussed in Section 5.2.

Common to all emotion-recognition work is the choice of which acoustic features to use in the analysis and what classification model to use. Features we chose fall into three groups: filter-bank energy features, frequency features, and voice-quality features as captured in periodicity, jitter, and shimmer. For our classification model we used the random forests method, which is well suited to emotion recognition tasks in which a large number of variables are derived from the acoustic features but there are relatively few observations.

In this paper we present results for the recognition of the four emotions Anger, Sadness, Joy, and Tension, plus Neutral. These are the emotions that are widely discussed in the literature on psychotherapy and shown in a particularly detailed analysis in Coan and Gottman (2007). Our results are reported for each couple separately. Although such speaker-dependent emotion recognition is rarely reported in the literature, there is good reason to believe that individual differences in the way emotion is expressed are important and that, as with speech recognition, the best results will be obtained when individual acoustic differences are taken into account.

We start by discussing in Section 2 other work that has been done on the recognition of emotions in naturally occurring speech. In Section 3 we describe our data collection and preparation, including the recording and transcription of the therapy sessions, the process we used for coding the emotions and associating emotions with speech, and the important issue of coder reliability, that is, the basis for claiming that our coding of emotions reflects objective scientific observation. In Section 4 we explain the acoustic features we used in our analysis and in Section 5 our classification model. Section 6 covers the challenge of unbalanced datasets. In Section 7 we present our results for the recognition of all the emotions plus Neutral and for pairwise recognition (Anger versus Sadness, for example). For the recognition of all the emotions plus Neutral, we also give results for each feature group separately. In Section 8 we discuss the significance of this study of emotion recognition and our approach to it.

## 2. Related work

Studies on the recognition of emotion in speech date back to the 1990s (Ververidis and Kotropoulos, 2003b,a; Dellaert et al., 1996; Nwe et al., 2001, 2003; Douglas-Cowie et al., 2003). A number of emotional speech datasets are now available in several languages (Ververidis and Kotropoulos, 2003b; Schuller et al., 2011). These include controlled datasets, with acted emotion and controlled speech, and natural datasets, with unrestricted speech and emotion. Controlled emotional speech datasets have attracted more attention in research because acted emotions permit a focus on specific emotions of interest. Acted emotions are somewhat exaggerated, however, and more intense than in naturally occurring situations. Acted emotional speech typically leads to higher recognition rates, but there are significant differences in the acoustic features that give the best recognition results for acted versus naturally occurring emotional speech (Vogt and André, 2005).

A number of natural emotional speech datasets now exist. The JST/CREST database has record- ings of domestic and social spoken interactions between volunteers throughout the day (Campbell, 2002). This very large dataset (> 1,000 hours) does not have any emotion coding, however, and can therefore not be used in its current state for automatic emotion recognition. Other natural emotional speech datasets consist of recordings from group discussions on emotive topics (Cowie et al., 1999; Douglas-Cowie et al., 2000) or TV shows and interviews. The Castaway database (Devillers et al., 2006) and the EmoTV corpus (Abrilian et al., 2005) are examples. A wide range of emotion labels were used in these datasets, which gives rise to difficulties in automatically recognizing emotions from the whole dataset.

To study the underlying processes in emotional speech with machine-based classification, emotions need to be narrowed down to a smaller subset. Such is the case in the datasets collected from call centers, therapy, and children's interaction with robots. In some cases, only two or three emotions were coded, such as negative versus non-negative in a call center with users interacting with a machine agent (Lee and Narayanan, 2003). Other datasets emphasize negative (frustrated) versus neutral in recordings from voice-controlled telephone services (Neiberg et al., 2006). Using data collected from AT&T's natural-language human-computer spoken dialog system, "How May I Help You?", seven emotions – positive/neutral, somewhat angry, somewhat frustrated, somewhat other negative, very angry, very frustrated, and very other negative were identified for automatic emotion recognition (Liscombe et al., 2005). The five emotion or attitudes of anger, excuse, fear, satisfaction, and neutral were noted in recordings of human-human interactions in a stock-exchange customer-service call center. Here joy and sadness were excluded because they are not common in that context (Devillers et al., 2002). In the CEMO corpus, which was obtained from a medical emergency call center, eight coarse-grained and 21 fine-grained emotions were labeled ( Devillers a nd Vasilescu, 2 006), although only the four emotions of anger, fear, relief and sadness were used in recognition. An important natural dataset is the FAU AIBO corpus, which features children playing with the Sony robot Aibo (Steidl, 2009). This corpus was adopted for the INTERSPEECH 2009 Emotion Challenge (Schuller et al., 2009), which featured the recognition of five emotions – anger, empathy, neutral, positive, and the rest – and the recognition of negative versus non-negative emotions. The natural emotional speech dataset that is closest to ours sought to identify patients who had major depression or were suicidal from spontaneous dialogue between patients and therapists (France et al., 2000).

Existing natural emotional speech datasets each have their own limitations. Some have a wide range of emotions, which creates difficulties for machine-learning models. Others have only a small number of emotions or several emotions dominated by negative or "other" emotions.

Higher recognition rates have, not surprisingly, been obtained on datasets with only two or three emotions. The best two-class recognition result achieved was 97.6% and it was for unbalanced datasets from call-center data (Lee and Narayanan, 2003). This work used a fuzzy inference classifier and 10 best features selected from 21 utterance-level summary statistic features. The best recognition rate for three emotions was 93% and it was achieved for the Swedish-language telephone service data using Gaussian Mixture Models (GMMs) over all frames of an utterance (Neiberg et al., 2006). For multiple-emotion recognition, an average recognition rate of 68% was obtained for five emotions using the stock-exchange dataset (Devillers et al., 2002). A balanced dataset was used for testing but not for training and lexical cues were included in the analysis. A recognition rate of 59.8% was achieved for four emotions in the CEMO corpus (Devillers and Vasilescu, 2006), with lexical cues again included in the analysis. Using the "How May I Help You" dataset and four groups of features – lexical, prosodic, dialog-act, and contextual – the best recognition rate achieved for seven emotions was 79%. However, 73.1% of the instances were labeled as non-negative in the dataset, producing a recognition baseline of 73.1% for random guessing (Liscombe et al., 2005).

## 3. Data collection and preparation

Our dataset of speech from couple-therapy sessions presents several advantages for data collection. Therapy sessions take place in an office where video and sound can be efficiently set up. Usually, participants are involved in enough sessions that emotions and emotion-word pairs that occur less frequently are not too infrequent over the course of all the sessions. More important, these therapy sessions are rich in expressed emotions in naturally occurring speech.

*3.1 Ethical review*

This research was approved by the Institutional Review Board of Stanford University as part of the study titled "Study of brain representations of language, emotion, visual images, music, and imagined stimuli using EEG recordings in individuals and couples." Reference number is IRB-11970. Written informed consent was obtained from each participant prior to his or her involvement in experimental sessions. Three professional psychotherapists were part of the research group.

*3.2 Recording and audio preprocessing*

The therapy sessions took place in a quiet room equipped for sound, video and EEG-brain recordings. The sessions consisted of a couple and a licensed psychotherapist and lasted about an hour (49 to 67 minutes). The male and female in the couple each wore an EEG geodesic sensor net with 128 sen- sors recording brain electrical activity. Three high fidelity Audio-Technica U873R microphones, placed on adjustable booms one foot from the head, recorded both participants and the therapist. A fourth microphone recorded the ambient sound. All microphones were connected to a 4-channel TASCAM DR-680 digital recorder digitizing the sound at a sampling rate of 44.1kHz at 24 bits per sample. A Sony Handycam HDR-XR160HD camcorder recorded a wide view of the couple, but not the therapist. The room was a lab at Stanford, set up only for this purpose, with divider screens separating from view most of the equipment. The participant couple sat on a couch facing the therapist who sat on a chair. The room was visually isolated from the outside and the noise level was low. The speech signal was re-sampled at 16 kHz in order to facilitate the analysis.

*3.3 Transcription*

The first major task after the recording was to transcribe the speech into text with precise onset times for each word. The performance of commercially available speech recognizers was too weak to be useful in our context. All transcriptions were performed manually. The transcription was faithful to the words and sounds actually used by the participants, without any grammatical or other corrections. Sounds such as coughing, laughing and other interjections were coded and transcribed accordingly. To achieve the timing accuracy needed for the analysis, we developed our own manual transcription software, written using MATLAB. In addition to listening to the recorded speech, the human transcriber also used a display of the spectrogram of the signal from each microphone to place the onset of each word within a 5-10 ms interval. This kind of transcription is extremely time- consuming, so for some of the sessions we transcribed only the segments where the emotional content was significant.

*3.4 Emotion codes*

Emotions can be coded according to categories, such as anger, joy or sadness, or dimensions such as arousal, valence, or control. For automatic emotion recognition in speech, there is a longer tradition of using a small set of categories. We chose the four emotions of Anger, Sadness, Joy, and Tension, along with Neutral, which refers to the absence of any of these four emotions. In this, we were guided by the efforts by others (Kerig and Baucom, 2004; Coan and Gottman, 2007; Ekman and Friesen, 1978) and, in particular, Gottman from whom we adopted our four emotions plus Neutral out of his 19 affective codes plus neutral. After careful examination of Gottman's 19 codes, we reduced the number to the four basic emotions mentioned above. The idea of basic emotions is not without controversy in the field of emotion research. Some theorists have their individual ideas of what the basic emotions are, others are more in agreement with each other, but this agreement is certainly not substantial (Ortony and Turner, 1990). It should be noted that the problem of what a specific word refers to, namely what counts as a basic emotion here, is not a problem unique to emotion theorists, but is found in ordinary language and in other parts of psychology. Almost everyone recognizes that there is something basic about fear, and also the three emotions of anger, sadness, and joy we chose. We included tension, as opposed to fear, in our set of emotions not because we have a strong position that tension is a basic emotion , but because something akin to fear occurs with high frequency in the interactions between members of a couple in psychotherapy and the notion of tension seems to capture it most closely. We recognize that the differences in opinion on this matter are real, and we are therefore treating our own selection as a tentative one for the purposes of moving research ahead.

*3.5 Coding procedure*

We developed our own software for the coding of the emotions to take advantage of the precise timings of the word onsets that our transcription offered. The program, written using MATLAB, allows the coder to watch the video recording of the couple while listening to the session, at the same time viewing the text transcript for each participant. The coder determines an emotion category and an intensity level (low, medium, high) of that emotion. (In the analysis reported in this paper, we did not differentiate between the intensity levels.) A coder estimates the time, $t_0$, at which an emotion begins, and the time, $t_1$, at which an emotion ends. Although data were recorded every millisecond, we did not expect the accuracy of $t_0$ or $t_1$, to be at this level.

The association of a word with an emotion code proceeds as follows for $C_i$ $\varepsilon$ {Anger,

Sadness, Joy, Tension, Neutral}. If at a time $t_n$ a coding is set for $C_i$ and at time $t_{n+1}$ a coding is set for emotion $C_j$ different from $C_i$, then any word with an onset in the interval $[t_n, t_{n+1}]$ is automatically coded as $C_i$, and any word with an onset immediately after $t_{n+1}$ is coded as $C_j$. We do not allow two emotions to overlap and every word occurrence (or token) is coded with one and only one emotion or Neutral. In the rest of this paper we talk about emotion-coded word tokens or just emotion-coded tokens. They refer to the segments of the acoustic signal associated with the word tokens and labeled with one of the four emotions or Neutral. Transformations of these segments are the observations that are used in the machine-learning classification model.

It is well recognized by most investigator that it is very expensive and time consuming to have the coding of the temporal length of emotion as an individual human coder's responsibility. The need for automated programming to do such coding is essential in the future to reduce cost, if nothing else (Cohn and Kanade, 2007; Bartlett et al., 2005).

*3.6 Coder reliability*

In studies using human judgment to label emotions or other occurrences that are the object of study, it is important to consider the basis for claiming that the labeling or coding reflects objective scientific observation. The coding procedure is presented in greater detail along with an analysis of coder reliability in (Crangle *et al.*, in preparation).

## 4. Acoustic features

There is a large body of literature on the production of speech in which acoustic features are defined according to a source-filter model (Hardcastle and Laver, 1999; Johnson, 2003). Most work on speech recognition and most studies of emotion in speech use this model as their basis. In this model, the source is created by the friction exerted by the air flow on the vocal cords, producing a sound wave. This source sound wave can be quasi-periodic when produce voicing sound or white noise for unvoiced sound. When this sound propagates forward, it resonates in the throat, mouth and nasal cavities, changing the spectral envelope of the sound. In the source-filter model, these cavities are modeled by a filter whose transfer function varies with the volume of these cavities. The speaker modulates the sound waves originating at the source by changing the shape and volume of the throat, mouth and nose cavities, and thus changing the filter transfer function. In natural speech, the source wave and the filter transfer function are approximately stationary when it is viewed in a short time interval of 10 to 40 ms. Thus most of the acoustic features of speech are defined for such time intervals called frames. In this study, the frame length is 40 ms and the shift between adjacent frames is 20 ms. The speaker controls speech production using muscles in the chest area for air-flow control, in the throat for vocal-cord tension, and in the mouth (tongue, jaw, lips) to shape the resonating cavities. The system is ultimately controlled by motor areas in the brain with feedback loops from the auditory and somatosensory systems (Pulvermüller et al., 2006; Guenther, 2006; Kent, 2000). Emotions can alter the control of these muscles and thus the acoustical features of the sound. These changes can be in amplitude, with an increase of the air flow corresponding to greater loudness, or in frequency, with a stronger tension of the vocal cord leading to a higher pitch. Some emotions also affect the variability of the features. As the hands of the man in anger may shake, the control of the different muscle groups involved in speech production may degrade to the point where some of the acoustical features are disturbed.

In our study, we use three groups of acoustic features to recognize emotions in speech. First are filter-bank energy features that describe the distribution of speech energy in the frequency domain. Second are frequency features. Third, we use voice-quality features to capture some of the voice- tremor qualities of emotional speech.

### 4.1 Filter-bank energy features

An auditory filter bank is a method that decomposes the speech signal by a series of band-pass filters, each corresponding to a perceptual frequency range. Several authors (Nwe et al., 2003; Ververidis and Kotropoulos, 2006) have shown that the energy of the speech, computed from the output of an auditory filter bank, can be used to recognize emotion in speech. Here we passed the continuous speech to a bank of 16 Gammatone filters (Patterson et al., 1 992) from 50Hz to 8000Hz, which was built using the MATLAB Auditory Toolbox (Slaney, 1998), and computed the 16 Log Frequency Power Coefficients (LFPC), defined by the mean energy in decibels of the output of each filter, normalized by its bandwidth. These we refer to as energy bands.

### 4.2 Frequency features

*Formants.* For a given configuration of the throat, mouth and nose cavities, the frequencies corresponding to peaks in the transfer function are amplified compared to other frequencies, corresponding to peaks in the sound spectrum. These concentrations of acoustic energy around particular frequencies in the speech sound wave are known as formants. The peaks define the formants and are numbered from lowest to highest frequency. The first two formants F1 and F2 mostly reflect the position of the tongue. The height of the tongue is related to F1 with a lower tongue position increasing F1. The horizontal position of the tongue is associated with F2 with a frontal position corresponding to a higher F2. These two features are most important for vowel identification, which are principally produced by the mouth-cavity modulation of the source quasi-periodic wave. F3 is found to reflect the rhoticity of speech sound (Broad and Wakita, 1977). The liquid consonant /r/ has very low F3. We estimate formants F1, F2 and F3 using the linear prediction method (LPC). The computation was performed with the COLEA software system (Loizou, 1998).

*Fundamental frequency F0.* Fundamental frequency F0 characterizes vibration frequency of the vocal cords, which can be perceived as pitch. Estimation of fundamental frequency is not simple and many methods have been proposed (Rabiner, 1977; Kadambe and Boudreaux-Bartels, 1992; Hess, 1982). We use the method of subharmonic-to-harmonic ratio (Sun, 2002) in this study.

### 4.3 Voice-quality features

*Periodicity.* Periodicity (Boersma, 1993; Thomson and Chengalvarayan, 1998) is a measure of how periodic the speech is in a frame. Periodicity is calculated as the maximum of the ratio between the auto-correlation function along the frame and the highest peak value when the delay $\tau$ varies from 40 to 242 samples, corresponding to the frequency range of 66 to 400 Hz, given the sampling rate of speech signal is 16K.

$$Per(n) = \max_{\tau \in [40,242]} R(\tau) \qquad (1)$$

where $R(\tau) = r(\tau)/ r(0)$ is the normalized auto-correlation with $r(\tau)$ being the auto-correlation function at $\tau$ and $r(0)$ the local maximum of the auto-correlation.

*Jitter.* Speech jitter (Thomson and Chengalvarayan, 1998) measures the variation of the estimated fundamental frequency from one frame to the next. It is defined by

$$Jitter(n) = \frac{|P_n - P_{n-1}|}{\frac{1}{N}\sum_{i=1}^{N} P_i} \qquad (2)$$

where $P_n$ is the estimated F0 at frame $n$.

*Shimmer.* Another acoustic phenomenon that has been related to the expression of emotion in speech is the variation of amplitude from one frame to the next, or shimmer (Greene, 1972). We estimated shimmer using the formula described in (Li et al., 2007).

$$Shimmer(n) = \frac{|A_n - A_{n-1}|}{\frac{1}{N}\sum_{i=1}^{N} A_i} \qquad (3)$$

where $A_i$ is the peak amplitude in frame $i$.

### 4.4 Acoustic feature space

The full set of acoustic features from the three groups can lead to a large number of variables when calculating the classification of a given acoustical wave form into the four emotions plus Neutral. Without further manipulation, the number of variables is the number of basic features (16 filter bank and 4 frequency) multiplied by the number of frames in a unit of analysis, which in our case is the word. We used segments of length 240 ms for our analysis, that is, acoustical wave-form intervals of 240 ms were associated with each emotion-coded token. (Section 5.2 discusses interval length in more detail.) A frame length of 40 and offset of 20 gave us 12 frames. With 16 filter-bank features and 4 frequency features for each frame and the means of periodicity, jitter and shimmer across all the frames, we have (16 + 4) x 12 + 3 for a total of 243 variables. Even selecting a smaller set of features would lead to a large number of variables. Two strategies address the issue. The simplest is to collapse the time dimension by creating summary statistics of each feature, such as mean, maximum, extrema, and variance, although multiplying these statistics again leaves us with a large number of variables. Another strategy relies on choosing recognition models that are able to handle a large number of variables with a limited number of observations. We chose this second strategy and return to this point in the next section.

### 4.5 Batch-effect adjustment

Many factors, such as the exact position of the microphones, may change from one session to the next. This unwanted variation between the data collected in one session (batch) and another is known as the batch effect and it must be corrected for when data from several sessions are concatenated. The main effect in our analysis is on the energy features of the acoustic signal, as we used the same sound equipment in all the sessions and we assume the frequency response of the microphones stayed the same in all the sessions.

In order to evaluate and adjust for the batch effect for data with $m$ batches each with $n_i$ samples within batch $i$ for $i = 1...m$, for $f = 1...F$ we used an empirical Bayes method (Johnson et al., 2007; Walker et al., 2008). The band energy can be modeled by a location and scale model as $Y_{ijf} = \alpha_f + X\beta_f + \gamma_{if} + \delta_{if}\varepsilon_{ijf}$, where $Y_{ijf}$ is the band energy observation $j$ in batch $i$ of feature $f$, $\alpha_f$ is the mean feature, $X$ is a design matrix coding the features of interest, and $\beta_f$ is the vector of regression coefficients corresponding to $X$. The error terms, $\varepsilon_{ijf}$, can be assumed to follow a normal distribution with expected value of zero and variance $\sigma^2$. The $\gamma_{if}$ and $\delta_{if}$ represent the additive and multiplicative batch effects of batch $i$ for feature $f$.

According to this method the correct model without the batch effects is given by

$$\gamma_{ijf}^* = Y_{ijf} - \widehat{\alpha}_f - X\widehat{\beta}_f - \widehat{\gamma}_{if}^*/\widehat{\delta}_{if}^* + \widehat{\alpha}_f + X\widehat{\beta}_f.$$

The estimation of the parameters

$$\widehat{\alpha}_f, \ \widehat{\beta}_f, \ \widehat{\gamma}_{if}^* \ \text{and} \ \widehat{\delta}_{if}^*$$

is implemented in the following steps.

Step 1. The first step is to standardize the energy bands in such a way that they have similar means and variances. With this procedure we reduce the effects of variation of the energy bands between different features that could compromise the estimation of the batch effect.

To standardize the data we first make an estimate of the model parameters

$$\alpha_f, \ \beta_f \ \text{and} \ \gamma_f$$

for each feature using least square constraining

$$\sum_i n_i \widehat{\gamma}_{if} = 0 \ \text{for} \ f = 1...F$$

The variance across all $N$ samples is given by

$$\widehat{\sigma}_f^2 = 1/N \sum_{ij} \left( Y_{ijf} - \widehat{\alpha}_f - X\widehat{\beta}_f - \widehat{\gamma}_{if} \right)^2$$

The standardized data are given by

$$Z_{ijf} = Y_{ijf} - \widehat{\alpha}_f - X\widehat{\beta}_f/\widehat{\sigma}_f.$$

Step 2. Assuming the additive batch has a normal distribution and that the multiplicative batch follows an inverse gamma distribution

$$\gamma_{if} \sim N\left(Y_i, \tau_i^2\right) \ \text{and} \ \delta_i^2 \sim \text{Inverse Gamma}\left(\lambda_i, \theta_i\right),$$

we estimate the parameters

$$\bar{Y}_i, \tau_i^2, \lambda_i, \theta_i$$

of these distributions from the samples by the method of moments.

After estimating the parameters these distributions are assumed as prior distributions, in a Bayesian inference approach, that allows us to determine the posterior distributions. With these posterior distributions we can make a final estimate for the batch effect parameters as the expected values of these posterior distributions.

Step 3. In this final step we make the data fit with the batch effects removed with the batch-effect estimators

$$\gamma_{if}^* \ \text{and} \ \delta_i^{2*} \gamma_{ijf}^* = \widehat{\sigma}_f/\delta_{ig}^* \left( Z_{ijf} - \widehat{\gamma}_{if}^* \right) + \widehat{\alpha}_f + X\widehat{\beta}_f.$$

The batch effect was significant for a large number of variables in all three couples and we adjusted all the energy-band variables using the R software package Bioconductor ComBat (Johnson et al., 2007). These adjusted values were used in all the analyses we report on in this paper.

## 5. Classification model for the recognition of the emotions

A classification model is a function that takes an observation **x** from a set of $M$ classes and gives a predicted class, in other words, a best estimate about which class **x** belongs to. (For our purposes here, an observation **x** is the acoustic feature vector associated with an emotion-coded token.)

Some models can also give an estimate of the probability of the observation belonging to that class. Classification models are in fact the composition of $M-1$ projection functions $h_1, h_2, ..., h_{M-1}$ with a threshold function $D$

$$\mathbf{x} \longrightarrow D(\begin{bmatrix} h_1(\mathbf{x}) \\ h_2(\mathbf{x}) \\ ... \\ h_{M-1}(\mathbf{x}) \end{bmatrix}) \longrightarrow j.$$

In linear models the projection functions are inner products of the observations **x** and a set of coefficients $h_i$ of the same length as **x**. We assume the intercept is coded in **x**. In a non-linear model, such as a support vector machine (SVM) with a non-linear kernal, the functions $h_i$ are non-linear functions of the vector **x**.

From the number $c_{ij}$ of test trials of class $i$ classified as class $j$ we build the confusion matrix $\mathbf{C} = (c_{ij})$. The classification rate $r$ is the sum of the diagonal divided by the number of test trials

$$r = \frac{\sum_i c_{ii}}{\sum_{ij} c_{ij}}.$$

In order to obtain an unbiased estimate of the classification rate, the optimization of the model and the estimation of the classification rate have to be done with two sets of independent observations, the training set for the optimization of the model and the test set for the estimation of the classification rate. There are two main approaches, each with many variants. In cross-validation, the observations are partitioned into $K$ sets and $K$ models are trained using each time the union of $K$-1 sets for training and the remaining set for test. Each set of the partition is a test set one and one time only. The classification rate is estimated as the mean of the $K$ classification rates for each model. Cross-validation does not guaranty an unbiased estimator of the predicted performance on new observations (Hastie et al., 2009). It has been shown that cross-validation can lead to upward or downward systematic bias. The main alternative to cross-validation is bootstrapping (Efron, 1979), where $K$ models are trained using sets sampled with replacement. For each model the test set is made of the observations not in the training set. In the most basic form of bootstrapping the combined estimate is the average of the estimates obtained with each bootstrap, but other methods exist such as majority vote.

Our choice of classification model had to take into account the fact that we had a large number of variables, 243 if all acoustic features were used. We chose the method of random forests, which is derived from decision trees and uses the random selection of feature variables to construct a collection of decision tress.

*5.1 Random forests classification model*

A random forest (Breiman, 2001) is a set of $K$ decision trees, each trained with a bootstrap sample from the dataset. A subset of $m$ variables is chosen randomly without replacement for each node of a tree, and the tree is built until there are no more variables to choose from, without any pruning. The algorithm is as follow. If $N$ is the total number of observations and $M$ the number of variables and $K$ the number of trees, for each of the $K$ trees we start by taking a bootstrap sample of size $n < N$. Then we build a decision tree by, at each node, selecting randomly a small subset of $m << M$ variables until exhausting all variables. In each node of the tree we fit the classification model with the $m$ variables and find the cutoff.

An estimator of the prediction error on new observations is computed by averaging the error rates obtained by using the trees to classify the out-of-bag (OOB) observations corresponding to the bootstraps used to train each tree. This is an unbiased estimator of the error rate we would observe by using the set of *K* trees to classify a new sample, using a majority vote of the *K* predictions.

In this paper the recognition rates we report for the random forests classification derive from the values of this estimator. As a by-product, the random forests method gives importance scores, providing a relative ranking of the features. Such scores are a rather classical measure, defined as the difference in performance between the full model and a model without the variable being scored.

Because the sample sizes in this study are very large, we get highly significant *p* values, mostly less than 10E-10, and consequently we do not report them individually.

*5.2 Interval length*

In this section we discuss our decision to use intervals of 240 ms for each emotion-coded token, that is, intervals of 240 ms to derive the variables used in classification. All intervals begin at a word's onset. In order to claim our recognition rates give us insights into the acoustical differences between emotions, on the face of it we want the intervals leading to the best recognition rates to include as few occurrences of the following word as possible.

The duration of a word token is defined as the interval between the onset of the word and the onset of the next word from the same speaker, as recorded in the transcript. Note that durations so defined include periods of silence between a speaker's successive words and between one turn of speech and another.

Of the 90,992 word tokens in total across all couples and all sessions, 58,552 (64%) lasted 240 ms or less. Figure 1 shows the distribution of the word tokens in terms of their durations. The 4% of the words that lasted longer than 1,000 ms are not shown.

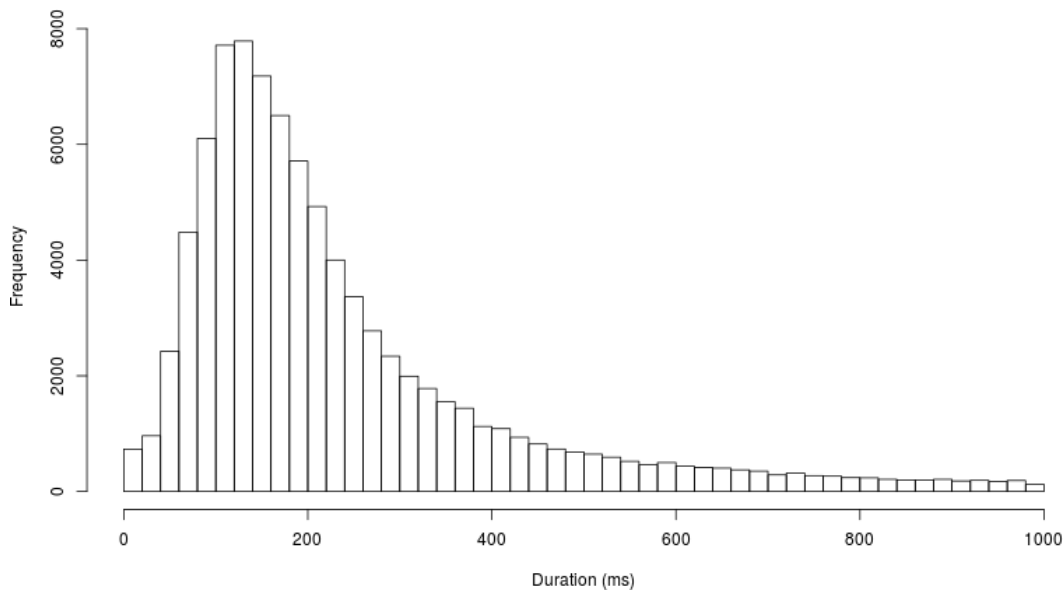

Figure 1: Histogram of the durations of the word tokens in all sessions for all couples

If the 240 ms interval is longer than a word's duration as defined it may include not only some silence but possibly also some or some portion of the words that follow. The intervals will, however, capture one and the same emotion across the interval. This follows from the way the emotions were coded and is consistent with the fact that many emotions, and in particular the four emotions we are interested in here, endure over periods longer than the duration of a word or several words, with some lasting beyond one or more whole utterances.

## 6. The challenge of unbalanced datasets

The crucial characteristic of a study of emotion in naturally occurring speech is that we do not have any control over how and when participants express emotions or what they say and when they say it. In a study without experimental control of the speech, choosing words to demarcate the units of analysis has the advantage of potentially providing a sufficient number of tokens of the same word type associated with at least some of the emotions. Obviously, we would have larger numbers at the syllable or at even lower levels but below the word level the segmentation becomes harder and less reliable. In a study of adults in therapy, there is no practical way to restrict the speech to a small set of phrases. Some idiomatic expressions such as "you know" may have a significant number of repetitions, but this is limited to one or two cases.

To understand the problem more clearly, we present details of the distribution of the words and emotions for our three couples across the 18 sessions. The first two sessions of Couple A had complete transcripts; all other sessions had partial transcripts, as described in section 3. Couple A had four sessions with a total of 18,664 words (tokens) and a vocabulary of 1,771 different words. That is, 18,664 word tokens were coded for one of the emotions or Neutral. The size of the vocabulary is the number of different words, also called *types*, that occur in and across all the sessions for a couple. It is important to note that "ah" and "ahh", for example, are different word types, as are "yes" and "yeah". For Couple B with 10 sessions, the total number of word tokens was 33,006 and the vocabulary size was 2,953. For Couple C the total number of word tokens was 17,746 and the vocabulary size was 1,620. Table 1 shows the number of emotion-coded word tokens or observations (including those for Neutral) in each session for the three couples. Table 2 shows the number of observations by emotion for the three couples. For each couple, the total across emotions matches the total number of word tokens in Table 1.

Table 1: Number of observations per session and vocabulary sizes for the couples

|  | 1 | 2 | 3 | 4 | 5 | 6 | 7 | 8 | 9 | 10 | Total | Vocab |
|---|---|---|---|---|---|---|---|---|---|---|---|---|
| Couple A | 10017 | 5671 | 442 | 2534 |  |  |  |  |  |  | 18664 | 1771 |
| Couple B | 2205 | 2205 | 2973 | 3448 | 4127 | 3375 | 2182 | 2860 | 5553 | 4078 | 33006 | 2953 |
| Couple C | 2316 | 4639 | 3751 | 7040 |  |  |  |  |  |  | 17746 | 1620 |

Table 2: Total number of observations by emotion for the couples

| Couple A | Joy | Sadness | Tension | Anger | Neutral |
|---|---|---|---|---|---|
| Female | 508 | 14 | 1304 | 174 | 3968 |
| Male | 915 | 755 | 1104 | 16 | 9906 |
| Couple B | Joy | Sadness | Tension | Anger | Neutral |
| Female | 3130 | 1712 | 8692 | 2025 | 3164 |
| Male | 2309 | 161 | 9033 | 551 | 2229 |
| Couple C | Joy | Sadness | Tension | Anger | Neutral |
| Female | 414 | 107 | 5497 | 2048 | 1248 |
| Male | 294 | 286 | 7168 | 48 | 636 |

These tables reveal the scope of the problem that results from an experiment with natural data. There are only 14 observations of Sadness coded for the female of Couple A and only 16 of Anger for the male, compared to large numbers for Tension and Neutral, for example. For the male of Couple C, there are only 48 observations for Anger. Furthermore, if we look at the size of the vocabulary compared to the total number of word tokens we see how sparse the coverage is, with many words occurring only once or a few times.

The consequences of the unbalanced distribution are two-fold. First, most classification models perform badly when the training set is significantly unbalanced. Second, unbalanced datasets, if not corrected for, compromise any inferences we may want to make about the intrinsic properties of the observations – in this case, the emotions as encoded in the acoustic properties. Suppose we have two classes and 10,000 instances we seek to classify. If our data consist of 9,900 instances of class 1 and 100 of class 2, a classifier that placed all 10,000 samples in class 1 would have a classification rate of 99%, but would tell us little if nothing about the classes and we could not make reliable inferences about the features used in classification.

There have been several methods proposed to deal with unbalanced datasets. The simplest and most effective when feasible is to constrain the sampling of the training set to be of equal or similar size in each class. The test set stays unbalanced, in which case this has to be accounted for when estimating statistical significance. Another method is to weight the observations, making it more expensive to misclassify observations in the smaller classes. A further approach is to over-sample the smaller classes by generating synthetic observations (Britsch et al., 2010; Chawla et al., 2002; Drummond and Holte, 2003). Combinations of these methods have also been used. A thorough study of the problem of unbalanced data, with several numerical examples, is to be found in the classic work of Jeffreys (1948). Here we chose a variation of the first method, as in most cases we do have enough observations in the smallest classes to make the second method work and we determined that the cost of the third method was not justified by the few cases where we had a very small sample in one class. In Chen et al. (2004) we can see the application of this sampling approach for unbalanced data using the random forests method.

For classifying the four emotions plus Neutral (a 5-class classification) we set the sample sizes in the training to be 80% of the observations in each class with a cap at a maximum of 500 observations. If a class had fewer than 100 observations, the class was dropped and we used a 4-class classification model instead. This approach reduced the extent of the imbalance in the training without imposing too rigid a constraint of strictly equal sizes. For the pairwise classifications (Anger versus Sadness, for example) we sampled a number equal to 80% of the smallest class for both classes. If a class had fewer than 200 observations, the class was dropped and that pairwise classification was not done.

## 7. Results

*7.1 Recognition of the four emotions plus Neutral*

The results for the recognition of the four emotions plus Neutral for all individuals are given in Table 3. Results are shown for each feature group independently and for all features taken together. The results are relative frequencies written for ease of notation as estimated percents. The last column shows which emotions plus Neutral were recognized.

For each individual, the best rates were achieved using the filter-bank features alone. It was also the case for each individual that the recognition rate for all features was somewhat lower than for filter-bank features but still better than the rate for frequency features alone, which was in turn better than for voice-quality features alone.

Table 3: Results for the recognition of the four emotions plus Neutral for female (F) and male (M) of Couples A, B, and C.

|          |   | FB  | F   | VQ  | ALL | No. observations | |
|----------|---|-----|-----|-----|-----|------------------|---|
| Couple A | F | 90% | 57% | 44% | 89% | 5954  | Joy, Tension, Anger, Neutral |
|          | M | 87% | 56% | 39% | 85% | 12680 | Joy, Sadness, Tension, Neutral |
| Couple B | F | 84% | 29% | 25% | 77% | 18723 | Joy, Sadness, Tension, Anger, Neutral |
|          | M | 81% | 44% | 33% | 78% | 14283 | Joy, Sadness, Tension, Anger, Neutral |
| Couple C | F | 88% | 45% | 37% | 85% | 9314  | Joy, Sadness, Tension, Anger, Neutral |
|          | M | 95% | 78% | 63% | 94% | 8384  | Joy, Sadness, Tension, Neutral |

FB is for filter-bank energy features, F for frequency features, VQ for voice-quality features, and ALL for all features.

Significant differences can be observed between individuals. For the female of Couple B, the recognition rate using frequency features alone was very low at 29% while for the male of Couple C, the rate using frequency features was 78%. Similarly, the rate using voice-quality features for the female of Couple B was even lower at 25% while it was 63% for the male of Couple C. The female of Couple A expressed very little Sadness across the recorded sessions (14 emotion-coded instances) and the male very little Anger (16 emotion-coded instances), with the consequence that Sadness for the female could not be included in the automatic recognition nor Anger for the male. In Couple C, the male also expressed too little Anger (48 emotion-coded instances) for it to be included in the automatic recognition.

These results suggest that individual differences in the frequencies of emotions expressed and in the features that provide the best recognition make speaker-dependent analysis necessary for the study of emotion expression in speech.

*7.2 Confusion matrices for the recognition of the four emotions plus Neutral*

The confusion matrices for the recognition of the emotions plus Neutral for each individual tell us more about the performance of our recognition model. Here in Tables 4, 5 and 6 we show the conditional probability matrices computed from the confusion matrices. They give for each matrix entry the conditional probability that an emotion, the one named in the row, was automatically recognized given that the emotion named in the column was the emotion being expressed in the judgment of the coders. The diagonal entries indicate the automatic classifications that were correct, that is, were aligned with the judgment of the emotion coders.

For Couple A, it is clear that we found it harder to classify the emotions of the female than those of the male. The evidence for this is in the diagonal entries. The recognition for Tension is similar for female and male (0.84 and 0.88) but for the other two emotions the results for the female (0.75 and 0.68) are lower than those for the male (0.87 and 0.82). web-based course(s) for the female are also more highly dispersed (range 0.25) than those for the male (range 0.06).

Table 4: Conditional probability matrices for Couple A.

Female

|  | Joy | Tension | Anger | Neutral | No. of Observations |
|---|---|---|---|---|---|
| Joy | **0.75** | 0.05 | 0.00 | 0.20 | 508 |
| Tension | 0.00 | **0.84** | 0.00 | 0.16 | 1304 |
| Anger | 0.01 | 0.09 | **0.68** | 0.22 | 174 |
| Neutral | 0.01 | 0.06 | 0.00 | **0.93** | 3968 |

Male

|  | Joy | Sadness | Tension | Neutral | No. of Observations |
|---|---|---|---|---|---|
| Joy | **0.87** | 0.03 | 0.01 | 0.09 | 915 |
| Sadness | 0.02 | **0.82** | 0.03 | 0.13 | 755 |
| Tension | 0.00 | 0.02 | **0.88** | 0.10 | 1104 |
| Neutral | 0.01 | 0.06 | 0.09 | **0.84** | 9906 |

In the case of Couple B, the conditional probability matrices for the male and female are more similar. The smallest diagonal entry for the classification of the four emotions plus Neutral is for the male and it is 0.65 for Sadness. For the female it is 0.73 for Tension. Dispersion is similar, with a range of 0.19 for the male and 0.17 for the female.

In the case of Couple C, for the classification of the four emotions plus Neutral, the smallest diagonal entry for the female is 0.74 for Sadness. For the male, the smallest diagonal entry for the classification of three emotions plus Neutral is 0.71 for Joy. The range for the female for four emotions is 0.14 while for the male it is 0.27 for three emotions. In general terms, there is probably less similarity between the male and female of Couple C than there is for the other two.

For all six individuals, the number of observations is quite large. Strong differences in the patterns of emotions between couples argue against speaker-independent emotion recognition and more specifically against using an out-of-sample method of recognition where training is on one couple and predictive recognition is on another.

Table 5: Conditional probability matrices for Couple B.

Female

|  | Joy | Sadness | Tension | Anger | Neutral | No. of Observations |
|---|---|---|---|---|---|---|
| Joy | **0.80** | 0.03 | 0.09 | 0.04 | 0.04 | 3130 |
| Sadness | 0.08 | **0.81** | 0.06 | 0.03 | 0.02 | 1712 |
| Tension | 0.12 | 0.05 | **0.73** | 0.06 | 0.04 | 8692 |
| Anger | 0.04 | 0.02 | 0.03 | **0.90** | 0.01 | 2025 |
| Neutral | 0.10 | 0.03 | 0.08 | 0.03 | **0.76** | 3164 |

Male

|  | Joy | Sadness | Tension | Anger | Neutral | No. of Observations |
|---|---|---|---|---|---|---|
| Joy | **0.80** | 0.00 | 0.14 | 0.00 | 0.06 | 2309 |
| Sadness | 0.07 | **0.65** | 0.22 | 0.02 | 0.04 | 161 |
| Tension | 0.11 | 0.00 | **0.78** | 0.01 | 0.10 | 9033 |
| Anger | 0.03 | 0.00 | 0.09 | **0.84** | 0.04 | 551 |
| Neutral | 0.08 | 0.00 | 0.14 | 0.01 | **0.77** | 2229 |

Table 6: Conditional probability matrices for Couple C.

Female

|  | Joy | Sadness | Tension | Anger | Neutral | No. of Observations |
|---|---|---|---|---|---|---|
| Joy | **0.77** | 0.00 | 0.10 | 0.10 | 0.03 | 414 |
| Sadness | 0.00 | **0.74** | 0.09 | 0.16 | 0.01 | 107 |
| Tension | 0.00 | 0.00 | **0.86** | 0.09 | 0.05 | 5497 |
| Anger | 0.00 | 0.00 | 0.08 | **0.88** | 0.04 | 2048 |
| Neutral | 0.01 | 0.00 | 0.09 | 0.06 | **0.84** | 1248 |

Male

|  | Joy | Sadness | Tension | Neutral | No. of Observations |
|---|---|---|---|---|---|
| Joy | **0.71** | 0.00 | 0.24 | 0.05 | 294 |
| Sadness | 0.00 | **0.79** | 0.19 | 0.02 | 286 |
| Tension | 0.00 | 0.00 | **0.98** | 0.02 | 7168 |
| Neutral | 0.01 | 0.01 | 0.23 | **0.75** | 636 |

*7.3 Pairwise recognition results*

The conditional probability matrices tell us about the probability of emotion *i* being recognized as emotion *j* among the five (or four) possible choices of the emotions plus Neutral. Pairwise recognition of emotion *i* versus emotion *j* gives us an estimation of the probability of recognizing an instance of emotion *i* as emotion *j* choosing only between these two emotions.

Figures 2, 3 and 4 show the pairwise recognition rates for the four emotions plus Neutral for Couples A, B and C. Figure 2 gives the results for the three emotions of Tension, Joy and Sadness, Figure 3 for Anger vs the other three emotions and Figure 4 Neutral versus the four emotions. Of the 74 recognition results computed, 47 were greater than or equal to 90%, 23 between 80% and 89%, and four below 80%. The highest recognition rate was 99% and it was for Tension vs Neutral in the female of Couple C. The lowest recognition rate was 76%, for Joy vs Tension and Neutral vs Tension in the male of Couple B. All sample sizes are given in Table 2.

Where there were too few observations (fewer than 200) for any emotion, its pairwise recognition was not computed. This occurred with Anger for the female of Couple A, Sadness for the male of Couple B, and Sadness for the female of Couple C. Note that Sadness for the female of Couple A, Anger for the male of Couple A, and Anger for the male of Couple C had already been excluded from automatic recognition because each had fewer than 100 observations.

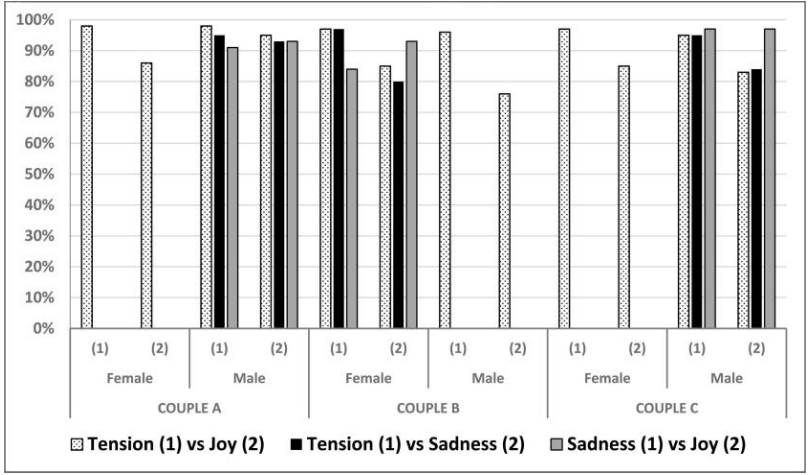

Figure 2: Pairwise Recognition rates for Tension, Sadness and Joy

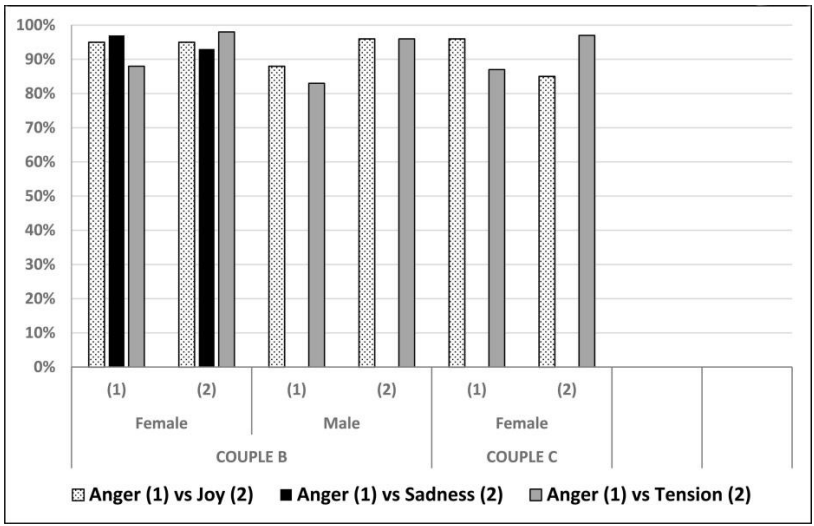

Figure 3: Pairwise recognition rates for Anger vs Joy, Sadness, and Tension

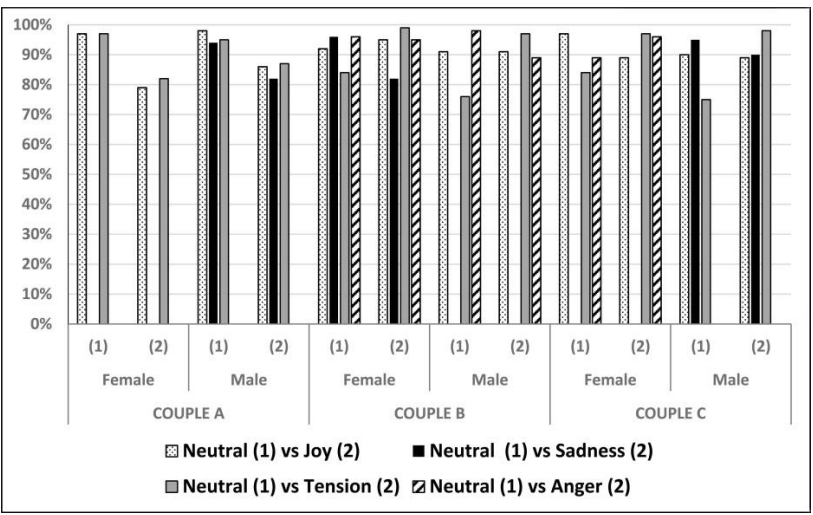

Figure 4: Pairwise recognition rates for Neutral vs Joy, Sadness, Tension, and Anger

# 8. Discussion

Our results show that couple therapy is a rich context for the study of emotions in naturally occurring speech. The recognition rates we achieved are significant enough that our results can be used to study the emotions of couples in psychotherapy without having to rely on time-consuming manual annotation of sessions. We conclude with a discussion of several key points.

*8.1 Speaker-dependent emotion recognition: the argument from theory*

We made the distinct choice to do single-speaker emotion recognition as opposed to multi-speaker emotion recognition. These two approaches, speaker-dependent versus speaker-independent, represent different conceptual and scientific models of emotion and speech. The important difference is that the kind of emotional speech found in psychotherapy is not, we would assert, ergodic.[2] To say that emotional speech is not ergodic is to say that the time averages of temporal analyses are not equal to the averaging over individual speaker data at a given point in time. Another way of characterizing the non-ergodic nature of emotional speech in psychotherapy is to point out that in non-ergodic processes the influence of the past does not fade away. It is widely recognized that the expression of emotion by clients in psychotherapy is very much a temporal process. Without consideration of this temporal processing, mistakes will be made in interpreting the results. (See Suppes (2002) for more details on the representation of such temporal data.)

In the history of psychology, B.F. Skinner objected to moving from the study of individuals to the study of groups and there is now a rich tradition in experimental psychology of single-subject study design, which owes much to his work (Skinner, 1953). In these Skinner-type studies, data are not averaged over subjects, but rather a large number of observations are collected over time from individual subjects and results are reported for each subject separately. We note that both the individuals and the couples in our study differ in interesting ways from each other. The emotions expressed by the two individuals in Couple A differed, with very little Sadness in the female and very little Anger in the male. In Couple B, both male and female expressed a fair amount of Joy (2309 instances for the female, 3130 for the male) and a great deal of Tension (9033 male, 8692 female) relative to the other two emotions. The female of Couple B did, however, also express Sadness and Anger far more than the male in the couple. For Couple C, the dominant emotion was Tension for both male (7168) and female (5497), and although the female also expressed a significant amount of Anger (2048), the male expressed too little Anger for that emotion to be included in the automatic recognition.

*8.2 The value of studying emotions in psychotherapy*

There are three aspects of the emotional speech we studied that make it particularly useful for the study of emotion. First, we have a natural speaker and a natural listener. In emotional speech datasets compiled from individuals interacting with a robot or with a machine agent in a call center, the absence of another person as the listener must surely have some effect on the emotional naturalness of the speech. Second, couples in psychotherapy typically participate in a number of sessions over the course of months or even years, providing data at various points over time. Such repeated measurements over time permit the study of single subject-emotion recognition and allow us to take account of the non-ergodic nature of emotions in psychotherapy. Finally, the nuance and subtlety of the emotions expressed in psychotherapy provide an unusual view of the interchange of emotions when two people (three with the therapist) are interacting.

---

[2] To be ergodic a process does not need to be a Markov process of some finite order. What is required is that the probabilistic influence of the past loses influence at what is termed in probability theory a geometrical rate. A good example would be stimulus-response models and learning theory, which in terms of observable random variables are chains of infinite order, not finite order, but under usual reinforcement schedules these processes are ergodic all the same (Lamperti and Suppes, 1959).

### 8.3 The unit of analysis in emotion recognition

An important topic to discuss is why we code emotions in relation to words. The answer is that in the kind of psychotherapy data we are collecting and analyzing, the occurrences of emotion are very much intertwined with the verbal expressions of emotion. Furthermore, these expressions of emotion in psychotherapy are subtle and often difficult to interpret. A coder or the therapist can have trouble identifying when an emotion begins, if no use is made of the verbal behavior occurring at the same time. It is difficult enough to know within a few milliseconds when a word begins. Except for rare cases, it is even more difficult to recognize when emotions begin. There is more to be said about this but we think from a practical standpoint this is the correct way to do the experiment at this time.

There is reason to believe from the various studies of emotion recognition that the optimal unit for emotion analysis is neither the word nor the longer conversational unit of a turn, but something of intermediate length. Automatically segmented intervals (called chunks) have been used. However, better results have been achieved with syntactic and semantically meaningful units as in Batliner et al. (2010), where sequences of words belonging to the same emotion class were taken as the unit of analysis. Our use of units that begin at a word's onset but have duration longer than the word in many instances means that we are using a syntactic and semantic demarcation while also allowing the capture of emotionally significant acoustic properties that last longer than the word. We do not use segments that differ in length from one observation to another, capturing the end of a word as well as its onset, because we chose to use not only summary statistics of the acoustic features in our analysis. That is, we chose not to collapse the time dimension but rather to take into account acoustic features in and across frames.

### 8.4 Future directions

There is plenty of room left for more detailed study of individual emotions. Deeper discussion and conception of whether or not there are a few basic emotions from which others can be constructed is not a settled question. In addition, like much experimental psychology, the theoretical framework for the recognition of results reported here is too static in character. The flow of speech and the flow of emotion are both among the most important examples of the temporal nature of much of human experience. Study of dynamic temporal processes is much more difficult both experimentally and theoretically, but in order to reach results of deeper scientific significance, such work is badly needed. This remark applies to both the fundamental theory and important applications. Even more pertinent from the standpoint of the main interest of this paper, this temporal quality of speech is matched very well by the temporal quality of emotions. The temporal flow of emotion probably has no natural grammar as is the case for speech or written language. This means that the study of emotion is more dependent on a thorough understanding of the ebb and flow of the emotions as a function of time. The complexity of such temporal study has necessarily delayed its deeper development. Fortunately, the wide-ranging nature of present research on emotion makes us hopeful that the temporal qualities of emotion will be more thoroughly studied in the near future.

# References

Abrilian, S., Devillers, L., Buisine, S., Martin, J.-C., 2005. EmoTV1: Annotation of real-life emotions for the specification of multimodal affective interfaces. In: 11th International Conference on Human- Computer Interaction (HCII). Las Vegas, NV.

Bartlett, M. S., Movellan, J. R., Littlewort, B. B., Frank, M. G., Sejnowski, T. J., 2005. Toward automatic recognition of spontaneous facial actions. In: Ekman, P., Rosenberg, E. (Eds.), What the Face Reveals. 2nd edition. Oxford University Press, Oxford, pp. 393–412.

Batliner, A., Seppi, D., Steidl, S., Schuller, B., 2010. Segmenting into adequate units for automatic recognition of emotion-related episodes: a speech-based approach. Advances in Human-Computer Interaction 2010, 15 pages, article ID 782802.

Boersma, P., 1993. Accurate short-term analysis of the fundamental frequency and the harmonics-to- noise ratio of a sampled sound. Proceedings of the Institute of Phonetic Sciences 17 (Proceedings 17), 97–110. URL http://www.mendeley.com/research/accurate-shortterm-analysis-of-the-fundamental-frequency-and-the-harmonicstonoise-ratio-of-a-sampled-sound-1/

Breiman, L., 2001. Random forests. Machine Learning 45 (1), 5–32. URL http://www.springerlink.com/index/U0P06167N6173512.pdf

Britsch, M., Gagunashvili, N., Schmelling, M., 2010. Classifying extremely imbalanced data sets. Electronics, 1–18. URL http://arxiv.org/abs/1011.6224

Broad, D. J., Wakita, H., 1977. Piecewise-planar representation of vowel formant frequencies. The Journal of the Acoustical Society of America 62 (6), 1467–1473.

Campbell, N., 2002. The recording of emotional speech. In: International Conference on Language Resources and Evaluation (LREC). Las Palmas, Spain, pp. 2029–2032.

Chawla, N. V., Bowyer, K. W., Hall, L. O., Kegelmeyer, W. P., 2002. SMOTE : Synthetic minority over-sampling technique. Journal of Artificial Intelligence Research 16 (1), 321–357. URL http://citeseerx.ist.psu.edu/viewdoc/summary?doi=10.1.1.18.5547

Chen, C., Liaw, A., Breiman, L., 2004. Using random forest to learn imbalanced data. URL http://digitalassets.lib.berkeley.edu/sdtr/ucb/text/666.pdf

Coan, J., Gottman, J., 2007. The specific affect coding system (SPAFF). In: Coan, J., Allen, J. (Eds.), Handbook of Emotion Elicitation and Assessment. Oxford University Press, Oxford, pp. 267–285.

Cohen, J., 1960. A coefficient of agreement for nominal scales. Educational and Psychological Measurement 20 (1), 37–46.

Cohn, J., Kanade, T., 2007. Use of automated facial image analysis for measurement of emotion expression. In: Coan, J., Allen, J. (Eds.), Handbook of emotion elicitation and assessment. Oxford University Press, Oxford, pp. 222–238.

Cowie, R., Douglas-Cowie, E., Romano, A., 1999. Changing emotional tone in dialogue and its prosodic correlates. In: ESCA Tutorial and Research Workshop on Dialogue and Prosody. Veldoven, The Netherlands, pp. 41–46.

Crangle, C.E., et al. (in preparation) The analysis of emotions in interacting couples during psychotherapy

Dellaert, F., Polzin, T., Waibel, A., October 1996. Recognizing emotion in speech. In: Spoken Language, Proceedings, International Conference on Spoken Language Processing (ICSLP96). Vol. 4. pp. 1970 – 1973, international Conference on Spoken Language Processing (ICSLP96). URL http://ieeexplore.ieee.org/xpls/abs_all.jsp?arnumber=608022

Devillers, L., Cowie, R., Martin, J.-C., Douglas-Cowie, E., Abrilian, S., McRorie, M., 2006. Real life emotions in French and English TV video clips: An integrated annotation protocol combining continous and discrete approaches. In: International Conference on Language Resources and Evaluation (LREC). Genova, Italy, pp. 1105–1110.

Devillers, L., Vasilescu, I., 2006. Real-Life Emotions Detection with Lexical and Paralinguistic Cues on Human-Human Call Center Dialogs. In: Proceedings of the Ninth International Conference Spoken Language Processing (ICSLP). pp. 801–804.

Devillers, L., Vasilescu, I., Lamel, L., 2002. Annotation and detection of emotion in a task-oriented human-human dialog corpus. In: ISLE Workshop on Dialogue Tagging. Edinburgh, United Kingdom.

Douglas-Cowie, E., Campbell, N., Cowie, R., Roach, P., Apr. 2003. Emotional speech: Towards a new generation of databases. Speech Communication 40 (1-2), 33–60. URL http://linkinghub.elsevier.com/retrieve/pii/S0167639302000705

Douglas-Cowie, E., Cowie, R., Schroder, M., 2000. A new emotion database: Considerations, sources and scope. In: ISCA Workshop on Speech and Emotion. Belfast, United Kingdom.

Drummond, C., Holte, R. C., 2003. C4. 5, class imbalance, and cost sensitivity: Why under-sampling beats over-sampling. In: Learning. Citeseer, pp. 1–8. URL http://citeseerx.ist.psu.edu/viewdoc/summary?doi=10.1.1.132.9672

Efron, B., 1979. Bootstrap methods: Another look at the jackknife. Annals of Statistics 7 (1), 1–26. URL http://projecteuclid.org/euclid.aos/1176344552

Ekman, P., Friesen, W., 1978. Facial Action Coding System: A technique for the measurement of facial movement. Consulting Psychologists Press, Palo Alto, CA.

France, D., Shiavi, R., Silverman, S., Silverman, M., Wilkes, D., 2000. Acoustical properties of speech as indicators of depression and suicidal risk. IEEE Trans Biomed Eng 47 (7), 829–837.

Greene, M. C. L., 1972. The voice and its disorders, 3rd Edition. Lippincott, Philadelphia.

Guenther, F. H., 2006. Cortical interactions underlying the production of speech sounds. Journal of Communication Disorders 39 (5), 350–365. URL http://www.ncbi.nlm.nih.gov/pubmed/16887139

Hardcastle, W. J., Laver, J., March 1999. The Handbook of Phonetic Sciences. In: Language. Vol. 75. Linguistic Society of America, pp. 152–154. URL http://www.jstor.org/stable/417486

Hastie, T., Tibshirani, R., Friedman, J., 2009. The Elements of Statistical Learning: Data Mining, Inference, and Prediction, 2nd Edition. Springer Series in Statistics. Springer, iSBN: 978-0-387- 84858-7. URL http://www-stat.stanford.edu/ tibs/ElemStatLearn/

Hess, W. J., 1982. Algorithms and devices for pitch determination of speech signals. Phonetica 39 (4-5), 219–240. URL http://www.karger.com/Journal/Issue/249729

Jeffreys, H., 1948. Theory of probability, 2nd Edition. Clarendon Press, Oxford.

Johnson, K., 2003. Acoustic and Auditory Phonetics, 2nd Edition. Wiley-Blackwell, iSBN 1-4051- 0122-

Johnson, W. E., Li, C., Rabinovic, A., Jan. 2007. Adjusting batch effects in microarray expression data using empirical Bayes methods. Biostatistics (Oxford, England) 8 (1), 118–27. URL http://www.ncbi.nlm.nih.gov/pubmed/16632515

Kadambe, S., Boudreaux-Bartels, F. G., 1992. Application of the Wavelet Transform for Pitch Detection of Speech Signals. IEEE Transactions on Information Theory 38 (2), 917–924. URL https://ieeexplore.ieee.org/stamp/stamp.jsp?tp=&arnumber=119752

Kent, R. D., 2000. Research on speech motor control and its disorders: a review and prospective. Journal of Communication Disorders 33 (5), 391–427; quiz 428. URL http://www.ncbi.nlm.nih.gov/pubmed/11081787

Kerig, P., Baucom, D., 2004. Couple Observational Coding Systems. Lawrence Erlbaum Associates, Inc. Publishers, Mahwah, NJ.

Lamperti, J., Suppes, P., 1959. Chains of infinite order and their application to learning theory. Pacific Journal of Mathematics 9 (3), 739–754.

Lee, C. M., Narayanan, S., 2003. Emotion recognition using a data-driven fuzzy interference system. In: European Conference on Speech and Language Processing (EUROSPEECH). Geneva, Switzerland, pp. 157–160.

Li, X., Tao, J., Johnson, M., Soltis, J., Savage, A., Leong, K., Newman, J., 2007. Stress and emotion classification using jitter and shimmer features. In: Acoustics, Speech and Signal Processing, IEEE International Conference on Acoustics, Speech, and Signal Processing (ICASSP). Vol. 4. pp. 1081–1084. URL http://ieeexplore.ieee.org/xpls/abs_all.jsp?arnumber=4218292

Liscombe, J., Riccardi, G., Hakkani-Tur, D., 2005. Using context to improve emotion detection in spo ken

dialog systems. In: International Conference on Speech and Language Processing (ICSLP). Lisbon, Portugal, pp. 1845–1848.

Loizou, P., 1998. COLEA: A MATLAB software tool for speech analysis. University of Texas at Dallas, Richardson, TX. URL http://www.utdallas.edu/ loizou/speech/manual.pdf

Neiberg, D., Elenius, K., Laskowski, K., 2006. Emotion recognition in spontaneous speech using GMMs. In: International Conference on Speech and Language Processing (ICSLP). Pittsburgh, USA, pp. 809–812.

Nwe, T. L., Foo, S. W., De Silva, L. C., 2003. Detection of stress and emotion in speech using traditional and FFT based log energy features. In: Fourth International Conference on Information Communications and Signal Processing 2003 and the Fourth Pacific Rim Conference on Multimedia Proceedings of the 2003 Joint. IEEE, pp. 1619–1623, iSBN 0-7803-8185-8. URL http://ieeexplore.ieee.org/xpl/articleDetails.jsp?arnumber=1292741

Nwe, T. L., Wei, F. S., De Silva, L., 2001. Speech based emotion classification. Electrical and Electronic Technology 2001 TENCON, Proceedings of IEEE Region 10 International Conference on Electrical and Electronic Technology 1, 297–301. URL http://ieeexplore.ieee.org/xpls/abs_all.jsp?arnumber=949600

Ortony, A., Turner, T., July 1990. What's basic about basic emotions? Psychol Rev. 97 (3), 315–331.

Patterson, R. D., Robinson, K., Holdsworth, J., McKeown, D., Zhang, C., Allerhand, M. H., 1992. Complex sounds and auditory images. In: Cazals, Y., Demany, L., Horner, K. a. (Eds.), Auditory Physiology and Perception. Oxford, Pergamon, pp. 429–446.

Pulvermüller, F., Huss, M., Kherif, F., Moscoso del Prado Martin, F., Hauk, O., Shtyrov, Y., 2006. Motor cortex maps articulatory features of speech sounds. Proceedings of the National Academy of Sciences of the United States of America 103 (20), 7865–7870. URL http://www.ncbi.nlm.nih.gov/pmc/articles/PMC1472536/

Rabiner, L., Feb 1977. On the use of autocorrelation analysis for pitch detection. IEEE Transactions on Acoustics, Speech, and Signal Processing 25 (1), 24–33. URL http://ieeexplore.ieee.org/lpdocs/epic03/wrapper.htm?arnumber=1162905

Schuller, B., Batliner, A., Steidl, S., Seppi, D., 2011. Recognising realistic emotions and affect in speech: State of the art and lessons learnt from the first challenge. Speech Communication 53 (9-10), 1062–1087. URL http://linkinghub.elsevier.com/retrieve/pii/S0167639311000185

Schuller, B., Steidl, S., Batliner, A., 2009. The INTERSPEECH 2009 emotion challenge. In: 10th Annual Conference of the International Speech Communication Association. pp. 312–315. URL http://www.isca-speech.org/archive/interspeech_2009/i09_0312.html

Skinner, B. F., 1953. cience and Human Behavior. Macmillan, New York.

Slaney, M., 1998. Auditory toolbox. Tech. Rep. 1998-010, Interval Research Corporation.

Steidl, S., 2009. Automatic classification of emotion-related user states in spontaneous children's speech. Logos Verlag, Berlin.

Sun, X., 2002. Pitch determination and voice quality analysis using Subharmonic-to-Harmonic Ratio. In: IEEE International Conference on Acoustics, Speech, and Signal Processing (ICASSP). Vol. 1. IEEE, pp. 333–336. URL http://ieeexplore.ieee.org/xpl/articleDetails.jsp?arnumber=5743722

Suppes, P., 2002. Representation and Invariance of Scientific Structures. CSLI Publications, Stanford, CA.

Suppes P, Nguyen MU, Nguyen DT, Nguyen C, Tuckner M, and Guimaraes, MP (2015). Suppes Brain Lab Psychotherapy EEG Dataset. Stanford Digital Repository. Available at: http://purl.stanford.edu/mz950kf4667

Thomson, D., Chengalvarayan, R., 1998. Use of periodicity and jitter as speech recognition features. In: Proceedings of the 1998 IEEE International Conference on Acoustics, Speech and Signal Processing, ICASSP '98 (Cat. No.98CH36181). Vol. 1. IEEE, pp. 21–24. URL http://www.mendeley.com/research/use-of-periodicity-and-jitter-as-speech-recognition- features/

Ververidis, D., Kotropoulos, C., 2003a. A Review of Emotional Speech Databases. In: Proceedings of the 9th Panhellenic Conference on Informatics PCI. No. November. Aristotle University of Thessaloniki, CiteSeer, pp. 560–574. URL http://citeseerx.ist.psu.edu/viewdoc/summary?doi=10.1.1.98.9202

Ververidis, D., Kotropoulos, C., 2003b. A state of the art review on emotional speech databases. In: Proceedings of the first R ichmedia C onference. pp. 109–119.

Ververidis, D., Kotropoulos, C., 2006. Emotional speech recognition: Resources, features, and methods.

Speech Communication 48 (9), 1162–1181. URL
http://linkinghub.elsevier.com/retrieve/pii/S0167639306000422

Vogt, T., André, E., 2005. Comparing feature sets for acted and spontaneous speech in view of automatic emotion recognition. In: ICME 2005. IEEE Computer Society, pp. 474–477.

Walker, W. L., Liao, I. H., Gilbert, D. L., Wong, B., Pollard, K. S., McCulloch, C. E., Lit, L., Sharp, F. R., 2008. Empirical Bayes accomodation of batch-effects in microarray data using identical repli- cate reference samples: application to RNA expression profiling o f b lood f rom D uchenne muscular dystrophy patients. BMC Genomics 9 (Supplement 2), 494–506.